\documentclass[conference]{IEEEtran}

\normalsize

\usepackage{graphicx}
\usepackage{epstopdf}

 \DeclareGraphicsExtensions{.pdf}

\usepackage[cmex10]{amsmath}
\usepackage{bm}
\usepackage{epsfig}
\usepackage{booktabs} 
\usepackage{latexsym}
\usepackage{enumerate}
\usepackage{multirow}
\usepackage{epstopdf}
\usepackage{longtable}
\usepackage{stfloats}
\usepackage{color} 
\usepackage{amssymb}
\graphicspath{{./Figures/}}
\usepackage{color}
\usepackage{tikz}
\usepackage{bbm}
\usepackage{authblk}
\usepackage{verbatim}
\usepackage{mathtools}

\usepackage[tight,footnotesize]{subfigure}
\usepackage{balance}

\usepackage{siunitx}
\DeclareSIUnit\bps{bps}
\DeclareSIUnit\Torr{Torr}
\DeclareSIUnit\torr{Torr}
\DeclareSIUnit\sample{Sa}
\usepackage{cite}

\usepackage{color, soul}
\newcommand{\tabincell}[2]{\begin{tabular}{@{}#1@{}}#2\end{tabular}}

\begin{document}

\title{Channel Measurement and Path-Loss Characterization for Low-Terahertz Indoor Scenarios} 
\author[1]{{Jia He}}
\author[2]{{Yi Chen}}
\author[2]{{Yiqin Wang}}
\author[1]{{Ziming Yu}}
\author[2]{{Chong Han}}

\affil[1]{
Huawei Technologies Co., Ltd, China. Email:  \{hejia83, yuziming\}@huawei.com}
\affil[2]{
Shanghai Jiao Tong University, China. Email:  \{yidoucy, wangyiqin, chong.han\}@sjtu.edu.cn}
\maketitle

\begin{abstract}
TeraHertz (THz) communications are envisioned as a promising technology, owing to its unprecedented multi-GHz bandwidth. In this paper, wideband channel measurement campaigns at 140 GHz and 220 GHz are conducted in indoor scenarios including a meeting room and an office room. Directional antennas are utilized and rotated for resolving the multi-path components (MPCs) in the angular domain. Comparable path loss values are achieved in the 140 and 220 GHz bands. To investigate the large-scale fading characteristics for indoor THz communications, single-band close-in path loss models are developed. To further analyze the dependency on the frequency, two multi-band path loss models are analyzed, i.e., alpha-beta-gamma (ABG) and multi-frequency CI model with a frequency-weighted path loss exponent (CIF), between which the ABG model demonstrates a better fit with the measured data. Moreover, a coherent beam combination that constructively sums the signal amplitudes from various arrival directions can significantly reduce the path loss, in contrast with a non-coherent beam combination.
\boldmath

\end{abstract}
\section{Introduction}
 New spectral bands are required to support Terabit-per-second (Tbps) data rates for future wireless applications to deal with the exponential growth of wireless data traffic~\cite{chong2017thz,chen2019channel}. Currently, wireless local area networks (WLAN) techniques, i.e., 802.11ad protocol and fifth-generation (5G) mobile networks, have opened up the millimeter-wave (mmWave) spectrum (10-100~GHz) to seek for broader bandwidth and higher data rates. However, still limited to several GHz bandwidth, the mmWave cannot support Tbps requirements. To further move up the carrier frequency, the Terahertz band spanning over 0.1 and 10~THz, is envisioned as one of the promising spectrum bands to enable ultra-broadband 6G communications.

The channel measurement is the fundamental of the channel studies at THz band. From the literature, a number of channel measurement campaigns at THz frequencies have been reported for indoor scenarios~\cite{yu2020wideband,yi2021Channel,xing2019indoor,kim2016characterization,eckhardt2019measurements,cheng2020thz,fu2020modeling,abbasi2020channel,song2020channel,serghiou2020ultra,nguyen2018comparing}. On one hand, indoor channel measurement campaigns focus on the short-range scenarios, e.g. on-desk, on computer motherboard, and inter-racks, and the distance ranges from 0.1m to 10m~\cite{kim2016characterization,eckhardt2019measurements,cheng2020thz,fu2020modeling,xing2019indoor,song2020channel}. On the other hand, the room-scale studies consist of very few transmitter-receiver (Tx-Rx) positions, generally less than 20 Tx-Rx positions due to the long time consumption of narrow beam scanning in the spatial domain~\cite{xing2019indoor,song2020channel,abbasi2020channel,serghiou2020ultra,nguyen2018comparing}. Therefore, an extensive channel study with various Tx and Rx positions in different indoor scenarios for different THz frequencies is still missing.  

\par In this paper, we first present a channel measurement campaign conducted in a meeting room and an office room at 140 GHz and 220 GHz, through frequency-domain channel sounding method via a vector network analyzer (VNA). The measured bandwidths are 13 GHz and 8 GHz at 140 GHz and 220 GHz, respectively. In particular, three cases, namely, Line-of-Sight (LoS) case in the office area, LoS case in the hallway and Non-Line-of-Sight (NLoS) case, are measured in an office room (not consistent). In light of the measurement results, we study the path loss properties at 140 GHz and 220 GHz in different indoor scenarios. The single-frequency path loss models as well as the multi-frequency path loss models are developed based on the channel measurement results. To be concrete, the omni-directional path loss, which considers the received power from all the scanned angles by the directional antenna, and the best-direction path loss, which only consider the received power from the strongest direction, are calculated and analyzed, respectively. In particular, different beam combination methods, i.e. coherent beam combination and non-coherent beam combination, are considered for the path loss calculation for the NLoS case.
\par The remainder of this paper is organized as follows. In Sec.~II, we describe the details of the THz channel measurement platform as well as the channel measurement campaign. Then, the single-frequency and multi-frequency path loss models are developed and derived with different post-processing methods based on the channel measurement results in Sec.~III. Finally, the paper is concluded in Sec.~IV.
\section{Channel Measurement Campaign}
In this section, we describe the THz measurement campaign, including the specification of the hardware system, indoor environments, and measurement deployment. Moreover, system calibration is carried out for eliminating the impact of the measurement system on the channel.

\subsection{Channel Measurement System at 140 GHz}
\par The THz channel measurement platform at 140~GHz consists of radio frequency (RF) fronts with horn antennas at both Tx and Rx sides, and a VNA. The local oscillator (LO) signal of 10.667~GHz is multiplied by a factor of 12 to 128 GHz. The immediate frequency (IF) signals generated by VNA range from 2 GHz to 15 GHz, which are mixed with the multiplied LO signal to the frequency band from 130 to 143~GHz. The measured bandwidth $B_w$ is 13 GHz. Therefore, the delay domain resolution of our measurement results, $\Delta t=1/B_w$, is 76.9~ps, which suggests two paths with the difference in propagation distance larger than 2.3~cm are resolvable. In addition, the number of the sampled points in the frequency domain or equivalently, the sweeping frequency points are 1301, which corresponds to the frequency interval of $\Delta f=10$~MHz. The maximum detectable delay, $\tau_m=1/\Delta f$, is calculated as 100 ns, hence, the largest traveling distance of a detectable path is $L_m=30$~m.

\begin{table}[htbp]
  \centering
  \caption{Parameters of the Measurement System.}
    \begin{tabular}{|l|c|c|}
    \hline
    \textbf{Parameter Value} & \multicolumn{2}{c|}{\textbf{Value}} \\
    \hline
    Sounder frequency & 140 GHz & 220 GHz \\
    \hline
    Local oscillator & 1.667 GHz & 18 GHz \\
    \hline
    Start frequency & 130 GHz & 201 GHz \\
    \hline
    End frequency & 143 GHz & 209 GHz \\
    \hline
    Bandwidth & 13 GHz & 8 GHz \\
    \hline
    IF bandwidth & \multicolumn{2}{c||}{10 MHz} \\
    \hline
    Sweeping points & 1301  & 801 \\
    \hline
    HPBW at Tx & 30$^{\circ}$   & 60$^{\circ}$ \\
    \hline
    HPBW at Rx & 10$^{\circ}$   & 10$^{\circ}$ \\
    \hline
    Delay resolution & 76.9 ps & 125 ps \\
    \hline
    Maximum excess delay & \multicolumn{2}{c|}{100 ns} \\
    \hline
    Maximum path length & \multicolumn{2}{c|}{30 m} \\
    \hline
    Azimuth rotation range & \multicolumn{2}{c|}{[0$^{\circ}$, 350$^{\circ}$]} \\
    \hline
    Elevation rotation range & \multicolumn{2}{c|}{[-20$^{\circ}$, 20$^{\circ}$]} \\
    \hline
    Rotation step & \multicolumn{2}{c|}{10$^{\circ}$} \\
    \hline
    \end{tabular}%
  \label{tab:mparameters}%
\end{table}%

\par  A directional horn antenna at Tx produces the half-power beamwidth (HPBW) of $30^\circ$ with an antenna gain of 15 dBi at 140~GHz, to guarantee a wide angular coverage. The Rx antenna gain is 25 dBi, and the HPBW is $10^\circ$, which is one-third of that at Tx for high spatial resolution. The Rx is mounted on a rotation unit, which can be rotated by step motors. In addition, the power of the test signal is 1~mW, and the noise floor of our THz measurement platform is $-120$ dBm (with antenna gain). The detailed parameters of the measurement system are summarized in Table~\ref{tab:mparameters}.

\begin{figure*}[htbp]
\centering
    \subfigure[Meeting room]{\includegraphics[width=0.31\textwidth]{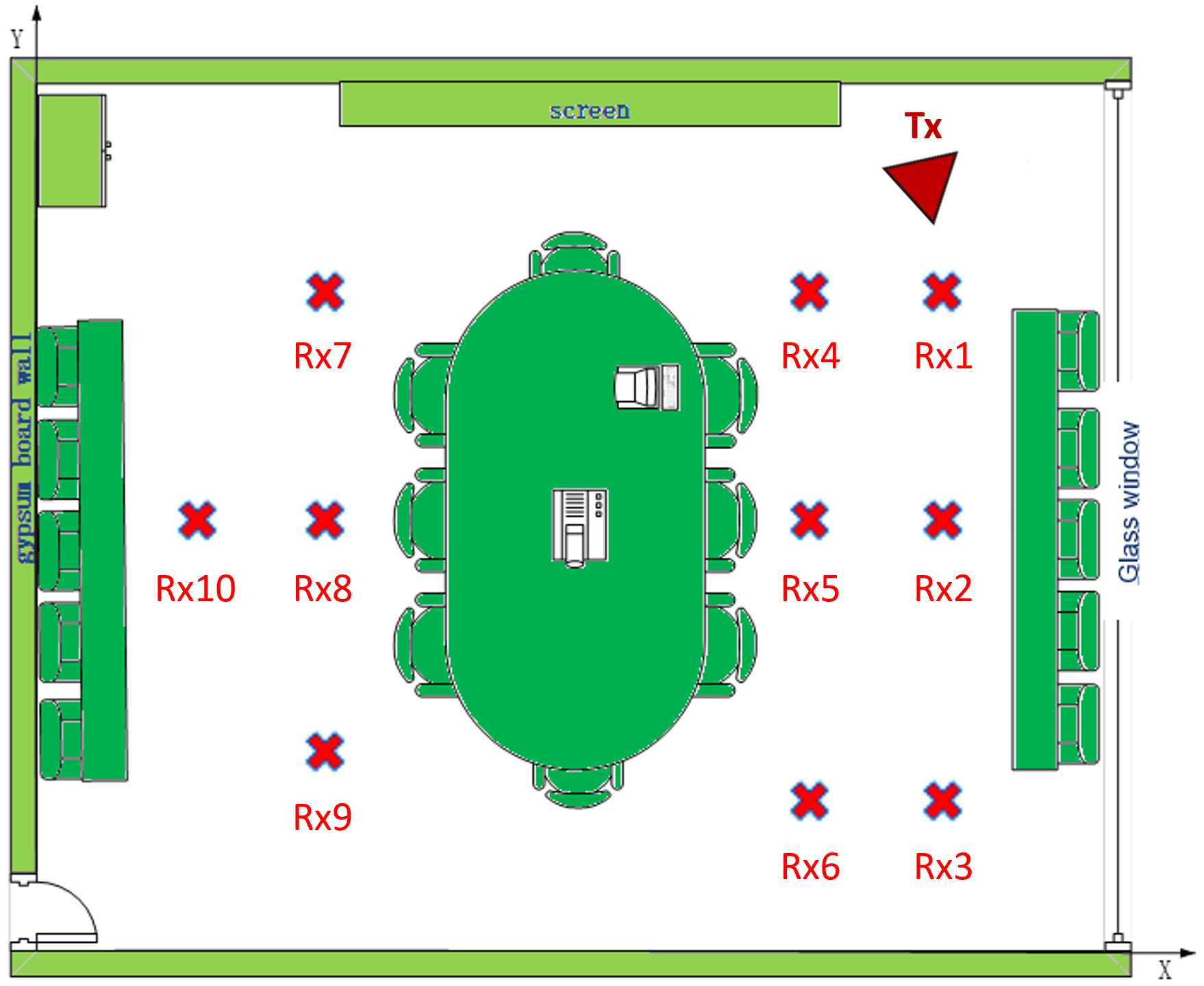}}
    \label{fig:deployment_meeting}
	\subfigure[Office room]{\includegraphics[width=0.65\textwidth]{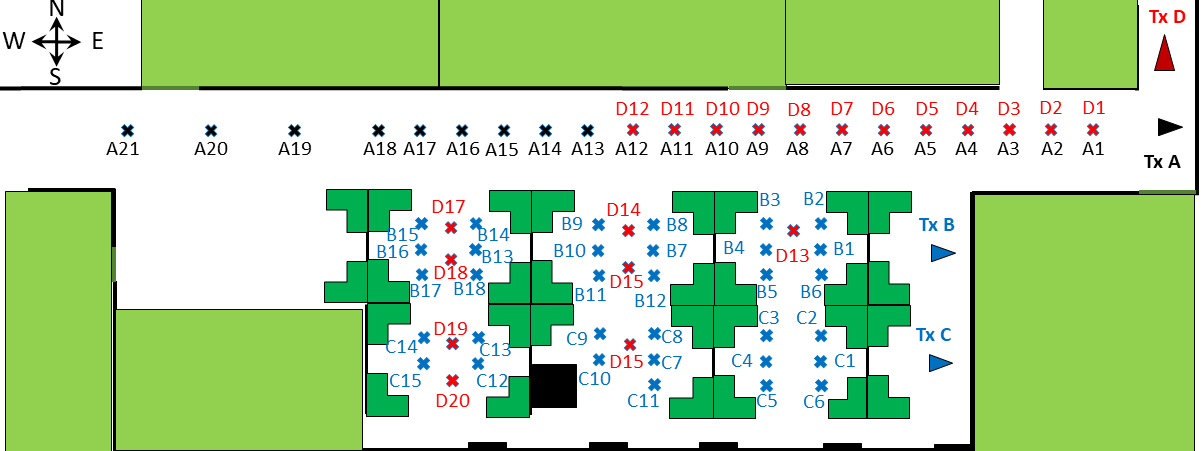}
	\label{fig:deployment_office}
	}
\caption{The deployment of the channel measurement in a (a) meeting room (b) office room.}
\label{fig:deployment}
\end{figure*}

\subsection{Channel Measurement System at 220 GHz}
\par The THz channel measurement platform at 220~GHz consists of RF fronts with horn antennas at both Tx and Rx sides and a VNA. The LO signal of 18~GHz is multiplied by a factor of 20 to 216 GHz. The IF signals generated by VNA range from 7 GHz to 15 GHz, which are mixed with the multiplied LO signal to the frequency band from 201 to 209~GHz. The measured bandwidth $B_w$ is 8~GHz. Therefore, the delay domain resolution of our measurement results is 125~ps. The HPBW of transmit antenna is $60^\circ$ while the HPBW of the receive antenna is $10^\circ$ at 220 GHz. The other parameters of the sounding system are summarized in Table~\ref{tab:mparameters} as well.


\subsection{Meeting Room Environment and Measurement Deployment}
We carry out the channel measurement in a typical meeting room with an area of 10.15 m $\times$ 7.9 m and a ceiling height of 4 m. In the meeting room, a 4.8 m $\times$ 1.9 m desk with a height of 0.77 m is placed in the center, and eight chairs are around the desk, as shown in Fig.~\ref{fig:deployment}(a). In addition, two TVs are closely placed in front of a wall. The material of one wall is glass, while the other three are lime walls. We notice that the maximum detectable path length imposed by the measurement system is 30~m, which is three times the dimension of the meeting room. As a result, reflected paths with at most three-order reflection can be recorded in our measurement. 
\par  In our measurement deployment, 10 positions of Rx are set in the meeting room,  as depicted in the top view of the meeting room in Fig.~\ref{fig:deployment}(a). Tx is close to a corner of the meeting room. In the measurement set, the Rx is placed on the positions Rx1-4 and Rx6-10. For the measurement of each Rx, the main beam of Tx is directed to the Rx. By contrast, Rx with the spatial resolution of $10^\circ$ scans the receiving beam in the azimuth domain from $0^\circ$ to $360^\circ$ and elevation domain from -$20^\circ$ to $20^\circ$ to detect sufficient multi-paths. As a directional beam of transmitter points to Rx, the antenna gain of the reflected paths from the ceiling and the floor are 16 dB lower than that of the LoS path at 140 GHz. Therefore, the considered reflected paths collected in our experiment are mainly from the desk, chairs and walls, whose elevation angles are sufficiently confined within [-$20^\circ$, $20^\circ$].

\subsection{Office Room Environment and Measurement Deployment}
The dimensions of the office room in our channel measurement campaign are 30~m $\times$ 20~m, as shown in Fig.~\ref{fig:deployment}(b), including a hallway and an office area. In the north of the office room, there is a 30-meter-long hallway. In the the office area, the space is partitioned by plastic boards into individual personal zones. On each desk, there are two monitors as well as other work-related items. 

\par The measurement campaign consists of three sets, (i) LoS office area, (ii) LoS hallway and (iii) NLoS, the deployments of which are depicted in Fig.~\ref{fig:deployment}(b). In the measurement set of LoS office area, Tx is placed at Tx B and Tx C, respectively. When Tx is placed at Tx B, Rx is placed at B1-B18. When Tx is place at Tx C, Rx is placed at C1-C15. There are in total 33 measurement points in the measurement set of LoS office area, and the distance between Tx and Rx varies from 3.5~m to 14~m. In the measurement set of LoS hallway, Tx is placed at Tx A while Rx is placed at A1-A21. The distance between Tx and Rx ranges from 2~m to 30~m. In the measurement set of NLoS, Tx is placed at Tx D, which is behind the corner of the hallway. 20 measured Rx points locate at D1-D20 without the existence of a LoS path, as shown in Fig.~\ref{fig:deployment}(b). The distance between Tx and Rx is 3.75~m-20~m. In the aforementioned measurement campaign, there are in total 74 measurement points. For each measurement point, the main lobe of Tx directs to the Rx in the LoS cases. In the NLoS case, Tx always directs to position A1. 

\subsection{System Calibration}
After channel measurements, system calibration need to be conducted to eliminate the effect of the VNA, cables and RF fronts at Tx and Rx. The process of system calibration requires to first measure the channel transfer function of an attenuator.
The measured S21 parameter from our channel measurement is $S_{\text{measured}}=H_{\text{system}}H_{\text{channel}}$ where $H_{\text{system}}$ is the response of the channel sounding system and $H_{\text{channel}}$ is the realistic channel transfer function of THz signals in indoor scenarios. Then, we connect the RF fronts at Tx and Rx with a attenuator and have the measured S21 parameter for calibration, $S_{\text{calibration}}=H_{\text{attenuator}}H_{\text{system}}$. Finally, the realistic channel transfer function of THz signals in indoor scenarios is represented as $H_{\text{channel}}=S_{\text{measured}}H_{\text{attenuator}}/S_{\text{calibration}}$.

\section{Path Loss Models} 
In this section, we first introduce the single-frequency path loss model, i.e., close-in (CI) model, and multi-frequency path loss, i.e., alpha-beta-gamma (ABG) model and multi-frequency CI model with a frequency-weighted path loss exponent (CIF). Then, we demonstrate the path loss from the channel measurement campaigns. The single-frequency and multi-frequency path loss models are developed and evaluated, respectively. In particular, the best-direction path loss and omni-directional path loss as well as path loss with different beam combination methods are considered. The properties and physical parameters revealed in this section are useful as guidelines for THz communication system design. 

 \begin{figure*}[htbp]

\centering
\subfigure[Path loss in the meeting room.]{
\includegraphics[width=0.48\textwidth]{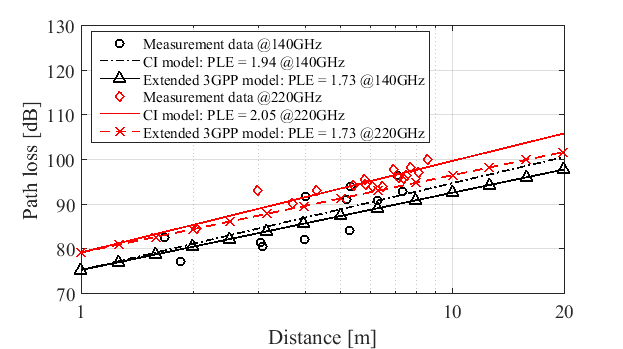}
\label{fig:pl_meeting}
}
\centering
\subfigure[Path loss in LoS office area case.]{
\includegraphics[width=0.48\textwidth]{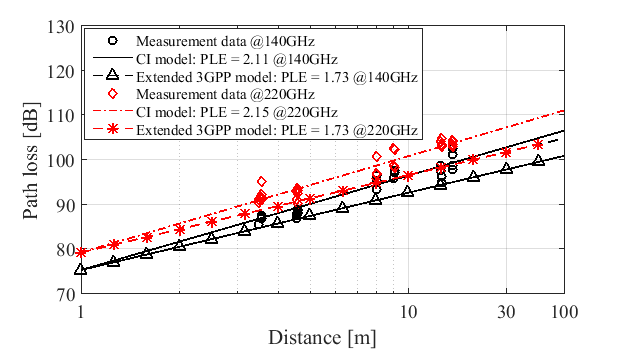}
    \label{fig:pl_office}
}

\centering
\subfigure[Path loss in LoS hallway case.]{
\includegraphics[width=0.48\textwidth]{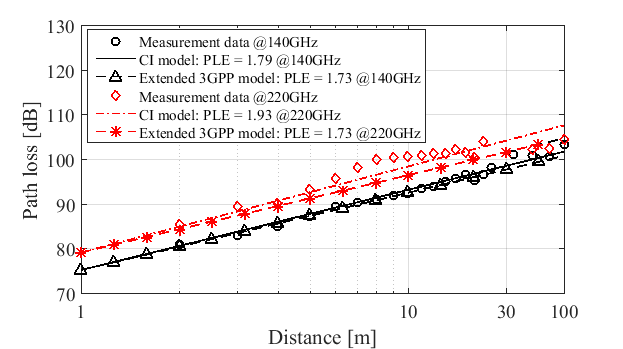}
\label{fig:pl_aisle}
}
\centering
\subfigure[Path loss in NLoS case.]{
\includegraphics[width=0.48\textwidth]{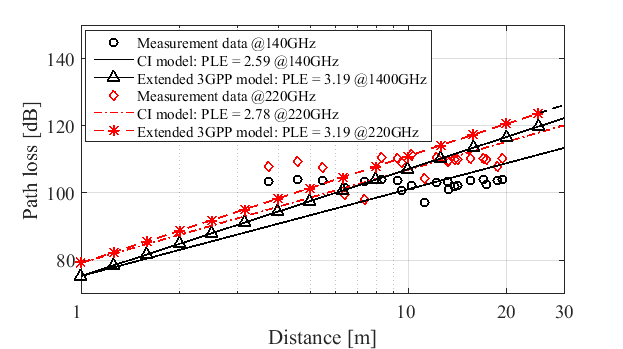}
\label{fig:pl_nlos}
}
\caption{Best-direction path loss measurement results and proposed CI model.}
\label{fig:pathloss} 
\end{figure*}
\subsection{Single-frequency Path Loss Model}
Path loss is a large-scale fading which reveals the signal power level of Rx at different places. We evaluate the close-in (CI) path loss model for all the measurement sets, respectively. In particular, the CI path loss model is represented as,
\begin{equation}
\text{PL}^{\text{CI}}[\text{dB}]=10\ \text{PLE}\ \log_{10}{(\frac{d}{d_0})}+\text{FSPL}(d_0)+X^{\text{CI}}_\sigma,
\label{eq:CI}
\end{equation}
where PLE is the path loss exponent, $d$ denotes the distance between Tx and Rx, and $d_0$ represents the reference distance which is 1~m in this work. $X^{\text{CI}}_\sigma$ is a zero-mean Gaussian random variable with standard deviation $\sigma^{\text{CI}}_{\text{SF}}$ in dB, which represents the fluctuation caused by shadow fading. Moreover, we compute the free-space path loss (FSPL) by invoking the Friis' law, given by
    \begin{equation}
    \text{FSPL}(d_0)=-20\log_{10}(\frac{c}{4\pi fd_0}),
    \label{eq:fspl}
    \end{equation}
where $c$ denotes the speed of light, and $f$ represents the carrier frequency. In addition, PLE in \eqref{eq:CI} is determined by minimizing $\sigma_{\text{SF}}$ via a minimum mean square error (MMSE) approach.

\subsection{Multi-frequency Path Loss Models}
Multi-frequency path loss models considers the dependency of path loss on both distance and frequencies, which requires the channel measurements at different frequencies in the same scenarios. There are two commonly used multi-frequency path loss models in the literature, i.e., ABG model and CIF model if antenna polarization is not considered as in this work.
\subsubsection{ABG model}
The alpha-beta-gamma (ABG) model is derived by adding a frequency-dependent optimization parameter to the current floating-intercept or alpha-beta (AB) model used in 3GPP~\cite{shu2016investigation}.
The ABG path loss model is represented as~\cite{shu2016investigation},
\begin{equation}
\text{PL}^{\text{ABG}}[\text{dB}]=10\alpha\log_{10}{(\frac{d}{d_0})}+\beta+10\gamma\log_{10}{(\frac{f}{f_0})}+X^{\text{ABG}}_\sigma,
\label{eq:ABG}
\end{equation}
where $f$ denotes the carrier frequency in gigahertz, and $f_0$ represent the reference frequency, which is 1~GHz in this work. $X^{\text{ABG}}_\sigma$ is a zero-mean Gaussian random variable with standard deviation $\sigma^{\text{ABG}}_{\text{SF}}$ in dB, which represents the fluctuation caused by shadow fading. The ABG model includes four model parameters, $\alpha$, $\beta$, $\gamma$, and $\sigma^{\text{ABG}}_{\text{SF}}$. $\beta$ is the offset value, while $\alpha$ and $\gamma$ show the dependence of path loss on distance $d$ and frequency $f$, respectively. Parameters are selected based on measured data by minimizing the value of $\sigma^{\text{ABG}}_{\text{SF}}$.

\subsubsection{CIF model}
To enable multi-frequency modeling, the CI model is generalized by adding a frequency-weighted path loss exponent. Therefore, the CIF path loss model is represented as~\cite{shu2016investigation},
\begin{equation}
\begin{split}
\text{PL}^{\text{CIF}}[\text{dB}]=10n\left(1+b\left(\frac{f-f_{\rm avg}}{f_{\rm avg}}\right)\right)\log_{10}{(\frac{d}{d_0})}\\
+\text{FSPL}(f,d_0)+X^{\text{CIF}}_\sigma,
\end{split}
\label{eq:CIF}
\end{equation}
where FSPL shares the same expression in~\eqref{eq:fspl}, and $X^{\text{CIF}}_\sigma$ is a zero-mean Gaussian random variable with standard deviation $\sigma^{\text{CIF}}_{\text{SF}}$ in dB.
Similar to the PLE in~\eqref{eq:CI}, $n$ measures the dependence of path loss on distance. The parameter $b$ measures the linear dependence of path loss on frequency about $f_{\rm avg}$, the weighted average of all frequencies, as,
\begin{equation}
f_{\rm avg} = \frac{\Sigma_{k=1}^{K}f_k N_k}{\Sigma_{k=1}^{K} N_k},
\label{eq:avg_freq}
\end{equation}
where $K$ is the number of frequencies considered in the CIF model, and $N_k$ is the number of data points corresponding to the $k$-th frequency $f_k$.

\subsection{Best-direction and Omni-direction Path Loss}
On one hand, in directional antenna channel measurement, the path loss can be divided by directional path loss and omni-directional path loss. In this paper, we consider the best-direction path loss as the directional path loss. That is, for each Tx-Rx position, we only calculate the path loss from the best direction which has the strongest received power. It should be noted that some researchers may consider another kind of directional path loss, i.e., they collect the path loss from all the scanned angles at each Tx-Rx position for the path loss regression model. The best-direction path loss for each Tx-Rx position is the path loss from the direction with the maximum received power scanned at Rx, which is calculated as,
\begin{equation}
    \text{PL}_{best}=-20*\log_{10}{(\max_{i,j}{{H^{\text{avg}}}_{i,j}})},
\end{equation}
where ${H^{\text{avg}}}_{i,j}$ is the channel transfer function (CTF) over the measured frequency band at $i^{\text{th}}$ azimuth angle and $j^{\text{th}}$ elevation angle at Rx. The calculation of ${H}_{i,j}$ is given as,
\begin{equation}
    {H^{\text{avg}}}_{i,j}=\sum_{s=1}^{S}H_{i,j,s},
\end{equation}
where $H_{i,j,s}$ is the CTF at $s^{\text{th}}$ swept frequencies and at $i^{\text{th}}$ azimuth angle and $j^{\text{th}}$ elevation angle at Rx. $S$ is the number of swept frequencies. The best-direction path loss results and the corresponding developed CI models for different indoor scenarios at 140 GHz and 220 GHz are shown in Fig.~\ref{fig:pathloss}, which are compared with the PLE for 0.5 to 100 GHz indoor channels given in 3GPP TR 38.901 model~\cite{3gpp2018study}.
\par On the other hand, the omni-directional path loss for each Tx-Rx position is the path loss considering the  power received from all the scanned angles at Rx, which is calculated as,
\begin{equation}
    \text{PL}_{omni}=-10*\log_{10}{(\sum_{i,j}{{H^{\text{avg}}}^2_{i,j}})}.
\end{equation}
\par The omni-directional path loss is generally lower than the best-directional as it involves all the received power. The PLE of CI models calculated with best-direction path loss and omni-direction path loss for indoor scenarios at 140 GHz and 220 GHz are summarized in Table~\ref{tab:PLE-CI}. It is observed that best-direction and omni-directional PLE values at 220 GHz are slightly higher than those at 140 GHz. As the best direction in LoS cases is the LoS direction and high-directional antenna are utilized at Rx, the best-direction propagation is approximately the free-space propagation. Therefore, the best-direction path loss in LoS cases is very close to 2, which is the PLE for free-space path loss. In addition, omni-directional PLE is around 0.5 higher than best-direction PLE, except in the NLoS case where the signal power is comparable to the noise floor. Therefore, omni-directional received power that sums the received power from all the directions will be much larger than the best-directional received power in NLoS case as noise power dominates the received power, which results in relatively significant difference between PLE with best-direction path loss and omni-directional path loss. By comparison, hallway scenario in the office room shows the lowest PLE and smaller than 2, which can be explained by the waveguide effect. Furthermore, PLE in office area is observed to be larger than PLE in meeting room. The reason is that the strong reflected paths from the walls are out of main beam of Tx in the office area due to the larger dimension of the office room than the meeting room~\cite{yi2021Channel}. 

\begin{table}[htbp]
  \centering
  \caption{PLE of CI models for indoor scenarios at 140 GHz and 220 GHz.}
    \begin{tabular}{|c|c|c|c|c|c|}
    \hline
       $f$ &  Path loss type  & Meeting & \tabincell{c}{Office\\area} & Hallway & NLoS \\
    \hline
    \multirow{2}{*}{\tabincell{c}{140\\GHz}} & Best-direction & 1.94  & 2.11  & 1.79  & 2.59 \\
\cline{2-6}          & Omni-directional & 1.44  & 1.67  & 1.25  & 1.78 \\
    \hline
    \multirow{2}{*}{\tabincell{c}{220\\GHz}} & Best-direction & 2.05  & 2.15  & 1.93  & 2.78 \\
\cline{2-6}          & Omni-directional & 1.61  & 1.72  & 1.36  & 1.99 \\
    \hline
    \end{tabular}%
  \label{tab:PLE-CI}%
\end{table}%
\par To further investigate the relationship among path loss, distance and frequencies for THz indoor communications, the multi-frequency path loss models, i.e., ABG model and CIF model, with best-direction path loss and omni-direction path loss are presented in Table~\ref{tab:MF-best} and~\ref{tab:MF-omni}, respectively. In ABG model for NLoS case, we observe that $\alpha$ is significantly lower than that in other scenarios, which suggests non-obvious linear dependency on the distance. The reason is that the received power is below the noise floor when distance is large. In addition, $\gamma$ in ABG model and $b$ in CIF model with both best-direction path loss and omni-direction path loss are all positive, which validates that the path loss is positively dependent on the frequency. Another observation is that the $n$ value of CIF models in a certain scenario is between the PLE values in CI models at 140 GHz and 220 GHz, respectively. The reason is that $n$ value represents the PLE at the reference frequency, $f_0$, which is the averaged measured frequencies in CIF models. This suggests that CI and CIF models have continuous relationship and offers the physical basis.
\par In order to evaluate the developed multi-frequency path loss models, we calculate the squired R values for each model. $\text{R}^2$ ranging from 0 to 1 indicates the goodness of fit of the model, e.g., $\text{R}^2=1$ indicates that the model predictions perfectly fit the data. Overall, ABG model shows higher $\text{R}^2$ values than CIF model, which suggests ABG model is more suitable for indoor THz channels.

\begin{table}[htbp]
  \centering
  \caption{ABG and CIF model with best-direction path loss}
    \begin{tabular}{|l|c|c|c|c|c|}
    \hline
    \multicolumn{6}{|c|}{ABG model with best-direction path loss}  \\
    \hline
          & \multicolumn{1}{c|}{$\alpha$} & \multicolumn{1}{c|}{$\beta$} & \multicolumn{1}{c|}{$\gamma$} & \multicolumn{1}{c|}{$\sigma_{\text{SF}}^{\text{ABG}}$ [dB]} & \multicolumn{1}{c|}{$\text{R}^2$}  \\
    \hline
    Meeting room & 2.21  & 21.65 & 2.41  & 2.80  & 0.82  \\
    \hline
    Office area & 2.17  & 28.31 & 2.17  & 1.74  & 0.91  \\
    \hline
    Hallway & 1.74  & 13.90 & 2.89  & 1.51  & 0.94  \\
    \hline
    NLoS  & 0.29  & 38.05 & 2.88  & 2.78  & 0.54  \\
    \hline
    \multicolumn{6}{|c|}{CIF model with best-direction path loss}  \\
    \hline
          & \multicolumn{1}{c|}{$n$} & \multicolumn{1}{c|}{$b$} & \multicolumn{1}{c|}{$f_0$ [GHz]} & \multicolumn{1}{c|}{$\sigma_{\text{SF}}^{\text{CIF}}$ [dB]} & \multicolumn{1}{c|}{$\text{R}^2$}  \\
    \hline
    Meeting room & 2.00  & 0.12  & 184.14 & 2.81  & 0.69  \\
    \hline
    Office area & 2.13  & 0.044 & 182.18 & 1.72  & 0.89  \\
    \hline
    Hallway & 1.86  & 0.16  & 178.00 & 1.64  & 0.93  \\
    \hline
    NLoS  & 2.68  & 0.16  & 180.00 & 5.71  & 0.50  \\
    \hline
    \end{tabular}%
  \label{tab:MF-best}%
\end{table}%

\begin{table}[htbp]
  \centering
  \caption{ABG and CIF model with omni-directional path loss}
    \begin{tabular}{|l|c|c|c|c|c|}
    \hline
    \multicolumn{6}{|c|}{ABG model with omni-directional path loss}  \\
    \hline
          & \multicolumn{1}{c|}{$\alpha$} & \multicolumn{1}{c|}{$\beta$} & \multicolumn{1}{c|}{$\gamma$} & \multicolumn{1}{c|}{$\sigma_{\text{SF}}^{\text{ABG}}$ [dB]} & \multicolumn{1}{c|}{$\text{R}^2$}  \\
    \hline
    Meeting room & 2.08  & 16.73 & 2.52  & 2.91  & 0.80  \\
    \hline
    Office area & 1.70  & 27.58 & 2.22  & 1.39  & 0.91  \\
    \hline
    Hallway & 1.29  & 11.54 & 2.94  & 1.67  & 0.90  \\
    \hline
    NLoS  & 0.067 & 27.27 & 3.09  & 1.19  & 0.88  \\
    \hline
    \multicolumn{6}{|c|}{CIF model with omni-directional path loss}  \\
    \hline
          & \multicolumn{1}{c|}{$n$} & \multicolumn{1}{c|}{$b$} & \multicolumn{1}{c|}{$f_0$ [GHz]} & \multicolumn{1}{c|}{$\sigma_{\text{SF}}^{\text{CIF}}$ [dB]} & \multicolumn{1}{c|}{$\text{R}^2$}  \\
    \hline
    Meeting room & 1.53  & 0.25  & 184.14 & 3.13  & 0.54  \\
    \hline
    Office area & 1.70  & 0.06  & 182.18 & 1.38  & 0.89  \\
    \hline
    Hallway & 1.30  & 0.19  & 178.00 & 1.80  & 0.84  \\
    \hline
    NLoS  & 1.88  & 0.25  & 180.00  & 3.98  & 0.52  \\
    \hline
    \end{tabular}%
  \label{tab:MF-omni}%
\end{table}%

\subsection{Path Loss with Beam Combination}
In THz communications, high-gain antennas or antenna arrays would be deployed to actively search and find the strongest directional beams especially when LoS path is obstructed. Combining the strongest beams can increase SNR and reduce path loss~\cite{rappaport2015wideband}. To be concrete, beams can be coherently combined and non-coherently combined. Coherent beam combination is to sum the amplitude of the signals from different directions. As a result, the path loss with coherently combining $N$ beams is calculated as,
\begin{equation}
    \text{PL}_{coherent}=-20*\log_{10}{(\sum_{k=1}^{N}{\overline{H}^{\text{avg}}_{k}})},
\end{equation}
where $\overline{H^{\text{avg}}}_{k}$ is the sorted averaged CTF over all the scanned directions at Rx and $\overline{H^{\text{avg}}}_{1}$ is the largest averaged CTF.
\par Similar to the omni-directional model procedure, non-coherent beam combination path loss consider the  of the power from different directions, which calculated as,
\begin{equation}
    \text{PL}_{non-coherent}=-10*\log_{10}{(\sum_{k=1}^{N}{{\overline{H}^{\text{avg}}}^2_{k}})}.
\end{equation}
\par The NLoS path loss with different beam combination methods and combined beams are summarized in Table~\ref{tab:NLOS-PL}. From Table~\ref{tab:NLOS-PL}, the best-drection path loss is equal to the path loss with only one beam combined. In addition, the coherent beam combination significantly reduces PLE compared with non-coherent beam combination in the NLoS case. With 5 strongest beams by coherent beam combination, PLE at 140 GHz in NLoS case reduces from 2.59 to 1.36. However, PLE is 2.01 at 140 GHz with 5 beams non-coherently combined. Although the path loss is reduced by coherent beam combination, we note that the noise power is enhanced by $N$ times after the beam combination. Moreover, We notice that the more beams are combined, the lower standard deviation of shadow fading in CI model is observed.

\begin{table}[htbp]
  \centering
  \caption{CI models with different beam combination methods}
    \begin{tabular}{|c|c|c|c|c|c|}
    \hline
          &       & \multicolumn{2}{c|}{140 GHz NLoS} & \multicolumn{2}{c|}{220 GHz NLoS}  \\
    \hline
          &       & PLE   & $\sigma_{\text{SF}}^{\text{CI}}$ [dB] & PLE   & $\sigma_{\text{SF}}^{\text{CI}}$ [dB]  \\
    \hline
    \multicolumn{2}{|c|}{Best direction} & 2.59  & 5.72  & 2.78  & 5.52  \\
    \hline
    \multirow{5}{*}{Coherent} & $N=1$     & 2.59  & 5.72  & 2.78  & 5.52  \\
\cline{2-6}          & $N=2$     & 2.05  & 4.60  & 2.53  & 4.54  \\
\cline{2-6}          & $N=3$     & 1.75  & 3.93  & 1.95  & 3.93  \\
\cline{2-6}          & $N=4$     & 1.53  & 3.46  & 1.74  & 3.50  \\
\cline{2-6}          & $N=5$     & 1.36  & 3.10  & 1.57  & 3.18  \\
    \hline
    \multirow{5}{*}{Non-coherent} & $N=1$     & 2.59  & 5.72  & 2.78  & 5.21  \\
\cline{2-6}          & $N=2$     & 2.34  & 5.16  & 2.53  & 5.03  \\
\cline{2-6}          & $N=3$     & 2.19  & 4.83  & 2.39  & 4.74  \\
\cline{2-6}          & $N=4$     & 2.09  & 4.60  & 2.29  & 4.54  \\
\cline{2-6}          & $N=5$     & 2.01  & 4.43  & 2.22  & 4.40  \\
    \hline
    \end{tabular}%
  \label{tab:NLOS-PL}%
\end{table}%
\section{Conclusion}
In this paper, we conduct channel measurement campaigns in indoor scenarios at 140 GHz and 220 GHz, respectively. The measured indoor scenarios includes a meeting room, and office area, hallway and NLoS in office room. Large-scale fading characteristics, i.e. path loss, in indoor scenarios at 140 GHz and 220 GHz are achieved and studied based on the channel measurement campaigns conducted in a meeting room and office room. The single-frequency and multi-frequency path loss models are developed and evaluated. From the analysis of the path loss models, we observe that the PLE in hallway scenario shows the lowest PLE among all the scenarios due to the waveguide effect. In addition, PLE in office area is higher than PLE in a meeting room, as the strong reflected paths from walls are not detected in the office area. Furthermore, the results show that PLE at 220 GHz is higher than that at 140 GHz, and this positive dependency of path loss on the frequencies is further validated by multi-frequency path loss models (ABG model and CIF model). The analysis of the $\text{R}^2$ values of the multi-frequency path loss models shows that ABG model outperforms CIF model in indoor THz channels. Furthermore, the coherent beam combination can significantly reduce the path loss in NLoS case.

\bibliographystyle{IEEEtran}
\bibliography{IEEEabrv,CY_bib,bibliography.bib}

\begin{thebibliography}{10}
\providecommand{\url}[1]{#1}
\csname url@samestyle\endcsname
\providecommand{\newblock}{\relax}
\providecommand{\bibinfo}[2]{#2}
\providecommand{\BIBentrySTDinterwordspacing}{\spaceskip=0pt\relax}
\providecommand{\BIBentryALTinterwordstretchfactor}{4}
\providecommand{\BIBentryALTinterwordspacing}{\spaceskip=\fontdimen2\font plus
\BIBentryALTinterwordstretchfactor\fontdimen3\font minus
  \fontdimen4\font\relax}
\providecommand{\BIBforeignlanguage}[2]{{%
\expandafter\ifx\csname l@#1\endcsname\relax
\typeout{** WARNING: IEEEtran.bst: No hyphenation pattern has been}%
\typeout{** loaded for the language `#1'. Using the pattern for}%
\typeout{** the default language instead.}%
\else
\language=\csname l@#1\endcsname
\fi
#2}}
\providecommand{\BIBdecl}{\relax}
\BIBdecl

\bibitem{chong2017thz}
C.~Han and Y.~Chen, ``{Propagation modeling for wireless communications in the
  terahertz band},'' \emph{IEEE Communications Magazine}, vol.~56, no.~6, pp.
  96--101, May 2018.

\bibitem{chen2019channel}
Y.~Chen and C.~Han, ``Channel modeling and characterization for wireless
  networks-on-chip communications in the millimeter wave and terahertz bands,''
  \emph{IEEE Transactions on Molecular, Biological and Multi-Scale
  Communications}, vol.~5, no.~1, pp. 30--43, Nov. 2019.

\bibitem{yu2020wideband}
Z.~Yu, Y.~Chen, G.~Wang, W.~Gao, and C.~Han, ``{Wideband channel measurements
  and temporal-spatial analysis for terahertz indoor communications},''
  \emph{in Proc. of IEEE International Conference on Communications Workshops
  (ICC Workshops)}, pp. 1--6, Jun. 2020.

\bibitem{yi2021Channel}
Y.~Chen, C.~Han, Z.~Yu, and G.~Wang, ``{140 GHz Channel Measurement and
  Characterization in an Office Room},'' \emph{in Proc. of IEEE International
  Conference on Communications (IEEE ICC)}, pp. 1--6, Jun. 2021.

\bibitem{xing2019indoor}
Y.~Xing, O.~Kanhere, S.~Ju, and T.~S. Rappaport, ``Indoor wireless channel
  properties at millimeter wave and sub-terahertz frequencies,'' \emph{in Proc.
  of IEEE Global Communications Conference (GLOBECOM)}, pp. 1--6, Dec. 2019.

\bibitem{kim2016characterization}
S.~Kim and A.~Zaji{\'c}, ``{Characterization of 300-GHz wireless channel on a
  computer motherboard},'' \emph{IEEE Transactions on Antennas and
  Propagation}, vol.~64, no.~12, pp. 5411--5423, Oct. 2016.

\bibitem{eckhardt2019measurements}
J.~M. Eckhardt, T.~Doeker, S.~Rey, and T.~K{\"u}rner, ``{Measurements in a real
  data centre at 300 GHz and recent results},'' \emph{in Proc. of 13th European
  Conference on Antennas and Propagation (EuCAP)}, pp. 1--5, Mar. 2019.

\bibitem{cheng2020thz}
C.-L. Cheng, S.~Sangodoyin, and A.~Zaji{\'c}, ``Thz cluster-based modeling and
  propagation characterization in a data center environment,'' \emph{IEEE
  Access}, vol.~8, pp. 56\,544--56\,558, Mar. 2020.

\bibitem{fu2020modeling}
J.~Fu, P.~Juyal, and A.~Zaji{\'c}, ``{Modeling of 300 GHz chip-to-chip Wireless
  Channels in metal enclosures},'' \emph{IEEE Transactions on Wireless
  Communications}, vol.~19, no.~5, pp. 3214--3227, Feb. 2020.

\bibitem{abbasi2020channel}
N.~A. Abbasi, A.~Hariharan, A.~M. Nair, and A.~F. Molisch, ``{Channel
  measurements and path loss modeling for indoor THz communication},'' \emph{in
  Proc. of 14th European Conference on Antennas and Propagation (EuCAP)}, pp.
  1--5, Mar. 2020.

\bibitem{song2020channel}
H.-J. Song, ``{LOS Channel Response Measurement at 300 GHz for Short-Range
  Wireless Communication},'' \emph{in Proc. of IEEE Wireless Communications and
  Networking Conference (WCNC) Workshops}, pp. 1--3, Apr. 2020.

\bibitem{serghiou2020ultra}
D.~Serghiou, M.~Khalily, S.~Johny, M.~Stanley, I.~Fatadin, T.~Brown, N.~Ridler,
  and R.~Tafazolli, ``{Ultra-Wideband terahertz channel propagation
  measurements from 500 to 750 GHz},'' \emph{in Proc. of International
  Conference on UK-China Emerging Technologies (UCET)}, Jul. 2020.

\bibitem{nguyen2018comparing}
S.~L. Nguyen, J.~Jarvelainen, A.~Karttunen, K.~Haneda, and J.~Putkonen,
  ``{Comparing radio propagation channels between 28 and 140 GHz bands in a
  shopping mall},'' 2018.

\bibitem{shu2016investigation}
S.~Sun, T.~Rappaport, T.~Thomas, A.~Ghosh, H.~Nguyen, I.~Kovacs,
  I.~Rodriguez~Larrad, O.~Koymen, and A.~Partyka, ``Investigation of prediction
  accuracy and parameter stability of large-scale propagation path loss models
  for 5g wireless communications,'' \emph{IEEE Transactions on Vehicular
  Technology}, vol.~65, pp. 1--1, 05 2016.

\bibitem{3gpp2018study}
{3GPP}, ``{TR 38.901: Study on channel model for frequencies from 0.5 to 100
  GHz (Release 16)},'' 3GPP Recommendation, Tech. Rep., 2020.

\bibitem{rappaport2015wideband}
T.~S. Rappaport, G.~R. MacCartney, M.~K. Samimi, and S.~Sun, ``Wideband
  millimeter-wave propagation measurements and channel models for future
  wireless communication system design,'' \emph{IEEE transactions on
  Communications}, vol.~63, no.~9, pp. 3029--3056, 2015.

\end{thebibliography}
\vfill

\end{document}